\documentclass[
reprint, 
 superscriptaddress,
 longbibliography,
 aps,
 prb,
]{revtex4-2}
\usepackage{caption}
\usepackage{graphicx}
\usepackage{dcolumn}
\usepackage{amsmath}
\usepackage{xspace}
\usepackage[T1]{fontenc}       
\usepackage{amssymb}
\usepackage{bm}
\usepackage[breaklinks=true,colorlinks,citecolor=blue,linkcolor=blue,urlcolor=blue]{hyperref}

\usepackage{tikz,xcolor,hyperref}
\usepackage[utf8]{inputenc}
\usepackage{makecell}
\usepackage[T1]{fontenc}
\usepackage{etoolbox}
\usepackage{txfonts}
\usepackage{xcolor}
\usepackage{fixltx2e}
\usepackage{ragged2e}

\makeatletter
\def\@email#1#2{%
 \endgroup
 \patchcmd{\titleblock@produce}
  {\frontmatter@RRAPformat}
  {\frontmatter@RRAPformat{\produce@RRAP{*#1\href{mailto:#2}{#2}}}\frontmatter@RRAPformat}
  {}{}
}%
\makeatother
\begin{document}

\preprint{AIP/123-QED}

\title{Dominant Role of Sulphur divacancy in Charge Trapping Dynamics in MoS$_2$}
\author{Srest Somay}
\affiliation{Department of Material Science and Engineering, Indian Institute of Technology Delhi, New
Delhi, 110016, India}
\author{Sitangshu Bhattacharya}
\affiliation{Electronic Structure Theory Group, Department of Electronics and Communication Engineering, Indian Institute of Information Technology Allahabad, Uttar Pradesh 211015, India}
\author{Krishna Balasubramanian}
\altaffiliation{All correspondence should be addressed to Dr. Krishna Balasubramanian, \href{}{bkrishna@mse.iitd.ac.in}}
\affiliation{Department of Material Science and Engineering, Indian Institute of Technology Delhi, New
Delhi, 110016, India}

\begin{abstract}
\noindent Intrinsic defects govern carrier trapping and recombination in two-dimensional semiconductors, yet the microscopic origin of defect-dependent capture dynamics remains unclear. Here, we compute carrier capture coefficients of vacancy defects, treating monolayer MoS$_2$ as a prototype, from first principles. We find that the single Sulphur vacancy forms a shallow defect with a small capture coefficient of $\sim 10^{-16}\ \mathrm{cm}^3/\mathrm{s}$, whereas the Sulphur divacancy exhibits a capture coefficient larger by seven orders of magnitude, $\sim 10^{-9}\ \mathrm{cm}^3/\mathrm{s}$, despite being only moderately deeper in energy. This enhancement originates from strong lattice relaxation enabling efficient multiphonon capture. Consequently, single vacancies contribute weakly to trapping, while Sulphur divacancies dominate nonradiative recombination and reduce quantum yield. In contrast, molybdenum vacancies and Sulphur antisites, although deep, show much smaller capture coefficients, indicating a limited role in carrier trapping in n-type devices.

\end{abstract}
\maketitle
\section{\label{sec:level1}INTRODUCTION}
\vspace{-0.45cm}
\noindent Two-dimensional transition metal dichalcogenides (TMDs) have emerged as promising channel materials for next-generation electronic devices owing to their atomically thin geometry and excellent electrostatic control. However, their practical deployment is often limited by intrinsic defects, which strongly influence carrier dynamics~\cite{Islam2026, das2021transistors, Ren2025, Zhou2013, Rasool2015}. A key manifestation is the pronounced hysteresis observed in the transfer characteristics of TMD field-effect transistors under cyclic gate-voltage sweeps~\cite{Karl2025, Kaushik2017, Dhosh2025, Park2016}. This behavior is widely attributed to carrier trapping and de-trapping at localized defect states. Crucially, only defects with trapping timescales comparable to the gate sweep rate can contribute to this effect. Identifying such defects, and thereby establishing the microscopic origin of hysteresis, remains an open and actively debated problem~\cite{Kaushik2017, Zhao2023, Shu2016}.\\
\noindent Several classes of traps can contribute to charge trapping in two-dimensional transistors, including intrinsic lattice defects (e.g., vacancies and grain boundaries), interface states at the dielectric/semiconductor boundary, and traps within the gate oxide. Recent theoretical studies based on the nonradiative multiphonon (NMP) framework suggest that oxide traps may dominate hysteresis, as their relaxation energies and defect level positions yield trapping timescales comparable to typical experimental sweep rates~\cite{Kaushik2017,Dhosh2025}. In contrast, intrinsic lattice defects are generally predicted to capture carriers on much shorter timescales, resulting in an essentially adiabatic response during gate-voltage sweeps\cite{Zhao2023, Dhosh2025}.\\
This discrepancy is particularly evident in molybdenum based TMD, such as MoS$_2$. Pronounced hysteresis has been observed even in suspended MoS$_2$ devices, where oxide traps are absent, pointing to an incomplete understanding of the role of intrinsic defects~\cite{Shu2016, Kaushik2017}. Experimental studies further highlight this complexity. Deep-level transient spectroscopy (DLTS) measurements have identified multiple Sulphur-related trap levels within the band gap, exhibiting widely varying capture coefficients~\cite{Ci2020, Zhao2023, Kim2022}. However, establishing a direct correspondence between these measured trap signatures and specific atomic-scale defect configurations remains challenging~\cite{Zhao2023, Ci2020}. This lack of microscopic assignment underscores the need for predictive, first-principles calculations of defect-specific capture coefficients.\\
\noindent First-principles implementations of the nonradiative multiphonon (NMP) framework have achieved quantitative accuracy in predicting carrier capture coefficients in conventional three-dimensional semiconductors~\cite{Alkauskas2014, Whalley2021, Turiansky2021}. Recent extensions to two-dimensional systems indicate that the formalism remains robust even in reduced dimensionality. For instance, calculations for monolayer hexagonal boron nitride predict exceptionally long nonradiative lifetimes ($>1$ ms) for vacancy-related transitions, consistent with their high radiative efficiency and experimentally observed single-photon emission~\cite{Wu2019, Chatterjee2025, Grosso2017}. However, a quantitative, defect-resolved understanding of carrier capture in TMDs is still lacking.\\
\noindent We provide a comprehensive first-principles analysis of electron capture by intrinsic vacancy defects in monolayer MoS$_2$ within the NMP framework. Focusing on the dominant defects in chemical vapor deposition-grown samples—Sulphur monovacancy, Sulphur divacancy, molybdenum vacancy, and Sulphur antisite—we identify a striking hierarchy in capture behavior. While both the monovacancy and divacancy introduce shallow levels near the conduction band minimum, their capture coefficients differ by seven orders of magnitude: $\sim 10^{-16}$ and $\sim 10^{-9}\ \mathrm{cm}^3\mathrm{s}^{-1}$, respectively. We demonstrate that this disparity cannot be explained by defect energetics alone. Instead, it originates from the magnitude of lattice relaxation, which controls the phonon overlap and thus the efficiency of multiphonon capture. The divacancy, despite being only slightly deeper, undergoes substantially larger structural distortion, leading to enhanced electron–phonon coupling and efficient carrier trapping. In contrast, the monovacancy exhibits weak lattice relaxation and correspondingly suppressed capture. Molybdenum vacancies and Sulphur antisites, although deep, are found to be ineffective traps due to large capture barriers.
\begin{figure*}[!ht]
  \centering
  \includegraphics[width=1.00\textwidth]{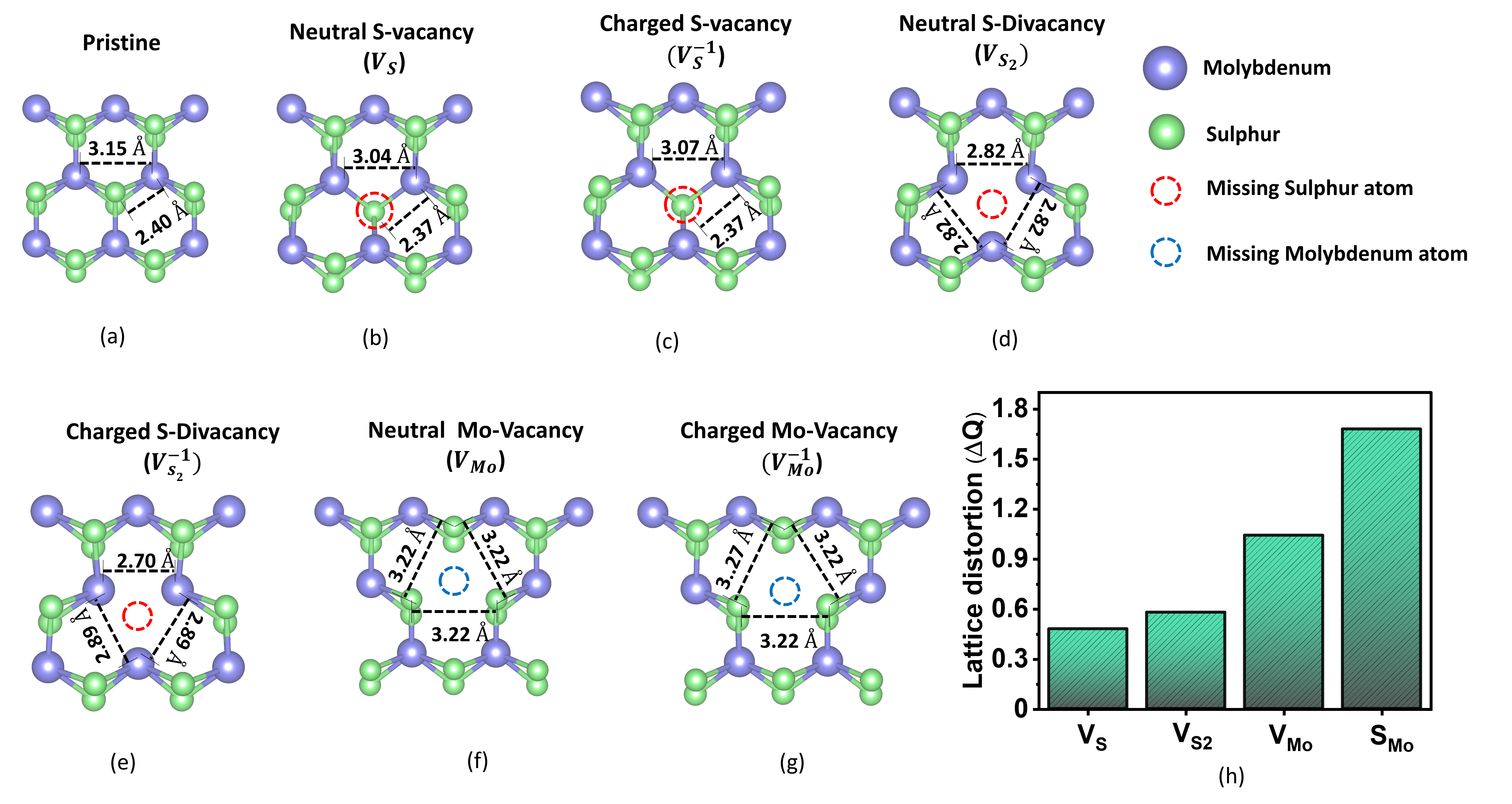}
  \caption{\justifying
Atomic structure of (a) Pristine MoS$_2$ (b) MoS$_2$ with neutral Sulphur monovacancy (c) Charged Sulphur monovacancy. (d) Neutral Sulphur divacancy (e) Charged Sulphur divacancy (f) Neutral Molybdenum vacancy (g) Charged Molybdenum vacancy. (h) lattice distortion associated with various intrinsic defects.}
  \label{fgr:Figure1}
\end{figure*}
\section{\label{sec:levela}COMPUTATIONAL METHODS}
\vspace{-0.45cm}
\noindent All first-principles calculations were performed using the Vienna Ab initio Simulation Package~\cite{PhysRevB.54.11169}. A hexagonal monolayer MoS$_2$ unit cell with lattice parameters $a=b=3.15$ Å was constructed based on previous studies~\cite{Komsa2015, Noh2014, Naik2018, Kim2022}. A vacuum spacing of 20 Å was introduced along the out-of-plane direction to eliminate interactions between periodic images. Vacancy defects were modeled using a $5\times5$ supercell. Structural relaxations were performed using the Perdew–Burke–Ernzerhof (PBE) exchange–correlation functional, as they are noted accurately predict the defect structure \cite{Komsa2015,Noh2014,PhysRevApplied.12.034038}. Atomic positions were optimized until the total energy converged below $10^{-9}$ eV using a $2\times2\times1$ Monkhorst–Pack $k$-point grid. Subsequent electronic structure calculations were carried out using the Heyd–Scuseria–Ernzerhof (HSE) hybrid functional with the same $k$-point sampling. The fraction of exact exchange was set to 0.30, yielding a band gap of $\sim 2.2$ eV for monolayer MoS$_2$. Charge transition levels of the defects were evaluated from total energies obtained within the HSE functional. Configuration coordinate diagrams were constructed by fitting defect formation energies along the structural distortion pathway within the harmonic approximation. Electron–phonon coupling relevant to carrier capture was computed at the \textbf{K}-valley by evaluating the derivative of the overlap between the conduction band minimum and defect states along the configuration coordinate. Phonon wavefunction overlaps and capture coefficients were calculated within the harmonic approximation using the Nonrad framework~\cite{Turiansky2021}. 
\begin{figure*}
\includegraphics[width=1.00\textwidth]{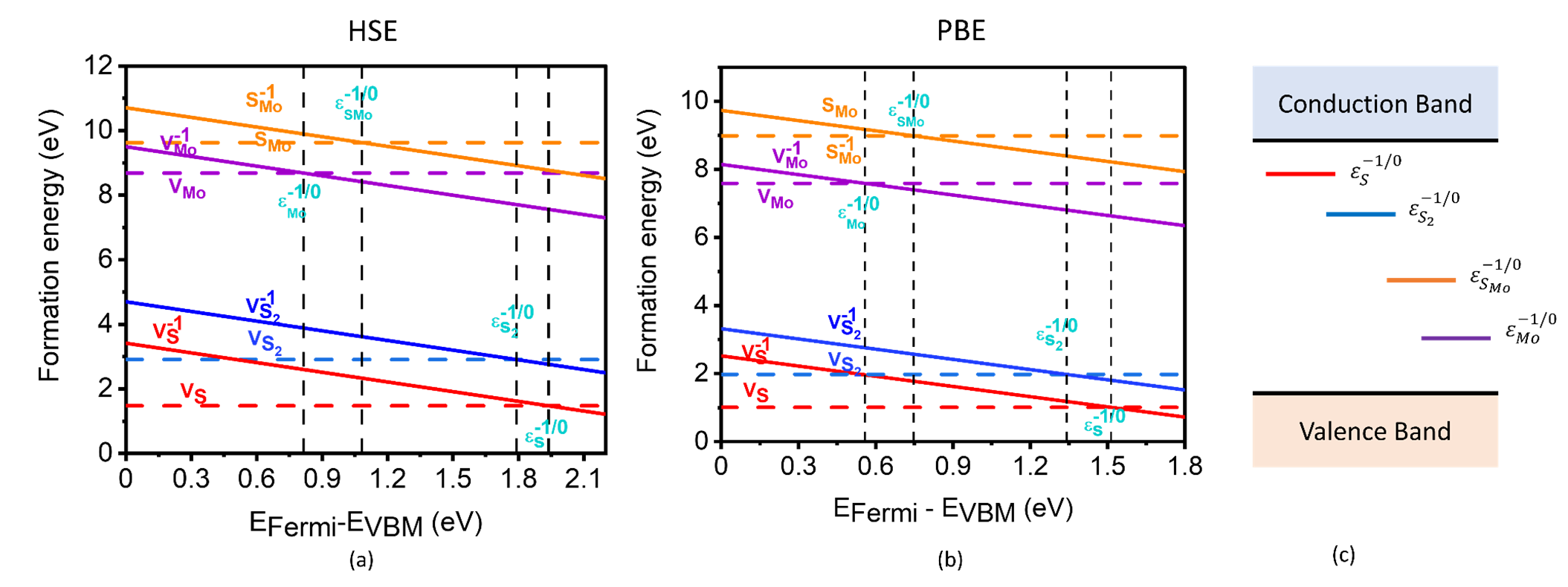}
  \caption{\justifying
Formation energy of neutral and charged Sulphur monovacancy, Sulphur divacancy, Molybdenum vacancy and Sulphur antisite using (a) HSE functional (b) PBE functional, (c) Charge transition level of various point defects showing that the  $V_s$ is closest to the conduction band and  $V_{Mo}$ is farthest.}
  \label{fgr:image2}
\end{figure*}
\section{RESULTS AND DISCUSSIONS}{\label{sec_3}}
\vspace{-0.45cm}
\noindent Figure \ref{fgr:Figure1}(a) shows the optimized structure of pristine monolayer MoS$_2$, which exhibits $D_3h$ symmetry. The equilibrium in-plane Mo–Mo distance is 3.15 Å, and the Mo–S bond length is 2.40 Å. The introduction of a neutral Sulphur vacancy ($V_S^0$) leads to a local inward relaxation of the neighboring atoms, as shown in Fig. \ref{fgr:Figure1}(b). This reconstruction reduces the in-plane Mo–Mo distance by 7.96$\%$ and the Mo–S bond length by 1.21$\%$, while preserving the local $C_3$ symmetry, consistent with previous reports~\cite{Noh2014, Jansen2024, Komsa2015}. Upon electron capture, the defect transitions to the negatively charged state ($V_S^{-1}$), shown in Fig. \ref{fgr:Figure1}(c). The additional structural relaxation is minimal: both the Mo–Mo and Mo–S bond lengths change only weakly relative to the neutral configuration. This indicates a small lattice response to charge capture for the Sulphur monovacancy. To quantify the structural distortion associated with carrier capture, we map the atomic displacements onto a one-dimensional configuration coordinate defined in terms of the mass-weighted displacement~\cite{Alkauskas2014}
\begin{equation}
\Delta Q^2 = \sum_{\alpha,t} m_{\alpha} \left( R_{i:\alpha t} - R_{f:\alpha t} \right)^2
\label{eqn:eqn1}
\end{equation}
Here, m$_\alpha$ is the atomic mass, R$_{i:\alpha t}$ and R$_{f:\alpha t}$ are the atomic coordinates of initial and final position. The structural response to carrier capture is quantified by the mass-weighted displacement $\Delta Q$, which measures the lattice distortion between the neutral and charged defect configurations. For the Sulphur monovacancy, we obtain a small distortion of $\Delta Q = 0.48\ \mathrm{amu}^{1/2}\text{\AA}$, consistent with the weak structural relaxation discussed above. Such small displacements are also observed in other 2D semiconductors \cite{Wu2019,PhysRevMaterials.9.026201,Lee2022}. In addition to the monovacancy, other intrinsic defects commonly observed in large-area grown MoS$_2$ include the Sulphur divacancy, Sulphur antisite, and molybdenum vacancy~\cite{Zhou2013,Rasool2015,Hong2015}.\\
A Sulphur divacancy ($V_{S_2}$) exhibits a substantially larger structural distortion compared to the monovacancy. In the neutral configuration, the defect undergoes significant bond contraction while preserving the local $C_3$ symmetry, indicating strong lattice strain. Upon electron capture, forming $V_{S_2}^{-1}$ (Fig. \ref{fgr:Figure1}(e)), this symmetry is broken. The neighboring Mo atoms form an isosceles triangle, with two Mo–Mo distances of 2.89 Å and one shortened bond of 2.70 Å. This symmetry breaking arises from a Jahn–Teller distortion, driven by the lifting of degeneracy in the Mo $d$-orbitals~\cite{Komsa2015,Noh2014,Jansen2024}. The associated lattice distortion is $\Delta Q = 1.68\ \mathrm{amu}^{1/2}\text{\AA}$, nearly four times larger than that of the Sulphur monovacancy. For comparison, the molybdenum vacancy ($V_{\mathrm{Mo}}$) leads to an outward relaxation of neighboring atoms due to the absence of the central atom. The charged configuration exhibits an asymmetric distortion, also consistent with a Jahn–Teller effect, with $\Delta Q = 1.04\ \mathrm{amu}^{1/2}\text{\AA}$ (Figs. \ref{fgr:Figure1}(f and g)). Similarly, the Sulphur antisite ($S_{\mathrm{Mo}}$) shows a relatively small distortion, $\Delta Q = 0.58\ \mathrm{amu}^{1/2}\text{\AA}$ (see Supplementary Information, SI \cite{Supplemental} for more details). A comparison of all defects (Fig. \ref{fgr:Figure1}(h)) reveals that the divacancy exhibits the largest lattice distortion, whereas the monovacancy and antisite show comparatively weak structural responses. This hierarchy of $\Delta Q$ directly reflects the strength of electron–phonon coupling and, as shown below, governs the corresponding carrier capture efficiencies.\\
\noindent To explore the charge trapping dynamics at the defect's sites in the monolayer, the formation energy of charged defect is an important parameter. The general formation energy of a defect (\(E_{f}^{0}\)) in the lattice is ~\cite{Freysoldt2014,Komsa2015}
\begin{equation}
E_{f}^{0} = E_{tot}^{0} - E_{tot}^{pristine} - \sum_{i}^{}{n_{i}\mu_{i}}
\end{equation}
where, \(E_{f}^{0}\) is the formation energy of defect in neutral state, \(E_{tot}^{0}\) is the total energy of structure with defect in neutral state. \(E_{tot}^{pristine}\) is the total energy of pristine structure, \(\mu_{i}\) is the chemical potential of the atom removed with \(n_{i}\) being its quantity. This energy level is represented with a dashed line in Fig. \ref{fgr:image2}(a). The charge state of a defect is governed by the position of the Fermi level, with electron capture occurring when the defect level becomes occupied. To analyze the stability of different charge states, we compute defect formation energies as a function of the Fermi level and construct charge transition level diagrams. In particular, we evaluate the formation energies of the defects in their negative charge states 
\begin{equation}
E_{f}^{q} = E_{tot}^{q} - E_{tot}^{pristine} - \left( \sum_{i}^{}{n_{i}\mu_{i}} \right)\  - q\left( E_{V} + E_{F} + \mathrm{\Delta}V_{0/p} \right)\
\label{eqn:eqn3}
\end{equation}
to determine the thermodynamic charge transition levels between different configurations~\cite{Komsa2015,Noh2014}. The last term in Eqn. (\ref{eqn:eqn3}) represents the electron chemical potential, where $q$ is the defect charge state, $E_V$ denotes the valence band maximum, and $E_F$ is the Fermi level, treated here as a free parameter. The term $\Delta V_{0/p}$ corresponds to the electrostatic potential alignment correction, obtained from the difference in the average electrostatic potential between the defective and pristine systems. The resulting formation energies are shown as solid lines in Fig. \ref{fgr:image2}(a), with each charge state represented by the same color as its corresponding neutral defect.
The defect formation energies are evaluated under Sulphur-poor conditions, with the Sulphur chemical potential obtained from the stability condition of MoS$_2$~\cite{Noh2014,Komsa2015}.
\begin{equation}
\mu_{s} = (\mu_{{Mos}_{2}} - \ \mu_{Mo(bcc)})/2\
\end{equation}
\begin{figure*}
\centering
  \includegraphics[width=1\textwidth]{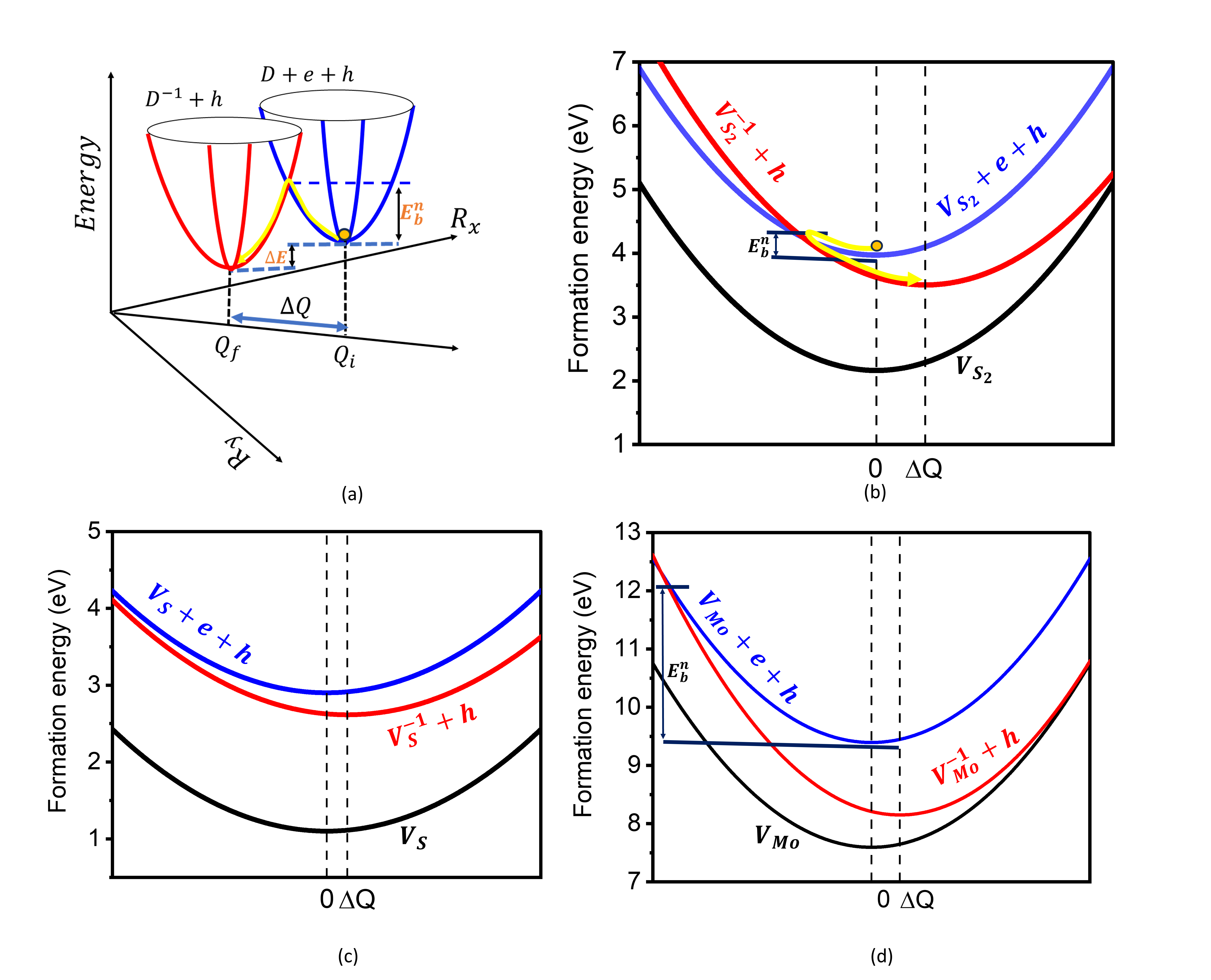}
    \caption{\justifying
(a) Schematic coordinate configuration diagram representing potential energy landscape for electron capture. Coordinate configuration diagram representing electron capture at (b) Sulphur divacancy (c) Sulphur monovacancy (c) Molybdenum vacancy (d) Sulphur antisite.}
  \label{fgr:Figure3}
\end{figure*}
Figure \ref{fgr:image2}(a) shows the charge transition level (CTL) diagram for the considered defects, computed using the HSE functional. For the Sulphur monovacancy ($V_S$), the neutral formation energy is 1.47 eV, in good agreement with previous theoretical reports~\cite{Komsa2015,Noh2014}. As expected, the formation energy of the neutral defect is independent of the Fermi level, whereas charged states exhibit a linear dependence with slope determined by the defect charge. The crossing point between different charge states defines the CTL. For $V_S$, the $\varepsilon^{-1/0}$ transition level is located at 1.94 eV above the valence band maximum (0.26 eV below the conduction band minimum), indicating a shallow defect level. This is consistent with both experimental observations and earlier theoretical studies~\cite{Komsa2015,Noh2014,Kim2022,Zhao2023,Ahn2023}. In contrast, the Sulphur divacancy ($V_{S_2}$) exhibits a significantly higher neutral formation energy of 2.90 eV, approximately twice that of the monovacancy. The charge transition level of the Sulphur divacancy ($V_{S_2}$) lies 0.40 eV below the conduction band minimum (CBM), slightly deeper than that of the monovacancy. In contrast, the molybdenum vacancy ($V_{\mathrm{Mo}}$) and Sulphur antisite ($S_{\mathrm{Mo}}$) exhibit much higher formation energies of 8.68 eV and 9.62 eV, respectively, indicating that these defects are less likely to form under Sulphur-poor growth conditions. Nevertheless, they are frequently observed experimentally, likely due to high-temperature growth processes. Their charge transition levels are located deep within the band gap, at 1.38 eV and 1.11 eV below the CBM for $V_{\mathrm{Mo}}$ and $S_{\mathrm{Mo}}$, respectively. Notably, the calculated transition levels for $V_S$ and $V_{S_2}$ are in close agreement with experimentally observed DLTS signals~\cite{Zhao2023,Ahn2023,Ci2020}, supporting the reliability of our approach. As summarized in Fig. \ref{fgr:image2}(c), both $V_S$ and $V_{S_2}$ introduce defect levels near the CBM, whereas $V_{\mathrm{Mo}}$ and $S_{\mathrm{Mo}}$ lie closer to the valence band maximum (VBM).\\
Accurate determination of charge transition levels requires a reliable description of the band gap, which can be well captured by the HSE functional\cite{Komsa2015,Noh2014}. However, due to its computational efficiency, we also evaluate CTLs using the PBE functional, as shown in Fig. \ref{fgr:image2}(b). For the Sulphur monovacancy ($V_S$) and divacancy ($V_{S_2}$), the CTLs are located 0.29 eV and 0.46 eV below the conduction band minimum (CBM), respectively, differing by only $\sim$30–60 meV from the HSE results. For the molybdenum vacancy ($V_{\mathrm{Mo}}$) and Sulphur antisite ($S_{\mathrm{Mo}}$), the CTLs are slightly deeper, with deviations of $\sim$100 meV compared to HSE, and are found at 1.24 eV and 1.05 eV below the CBM. Overall, the CTLs obtained from PBE and HSE are in close agreement for all defects, indicating that the relative defect level positions are robust with respect to the choice of exchange–correlation functional. While CTLs are independent of absolute formation energies, we additionally evaluate formation energies under Sulphur-rich conditions, with results provided in the SI~\cite{Supplemental}.\\
To evaluate the carrier capture coefficient, the process is described within the potential energy surface (PES) formalism. Figure ~\ref{fgr:Figure3}(a) illustrates a schematic configuration coordinate diagram. The initial state corresponds to a neutral defect with an electron in the conduction band and equilibrium atomic configuration $Q_i$, represented by the blue curve. Upon electron capture, the system transitions to a charged defect state with equilibrium configuration $Q_f$, shown by the red curve, which is lower in energy by $\Delta E$ (corresponding to the charge transition level). The transition from $Q_i$ to $Q_f$ involves a change in both electronic and atomic configurations and can be viewed as motion along the configuration coordinate connecting the two minima. The capture process is governed by the overlap between the initial and final states and is strongly influenced by the activation barrier $E_b^n$, with the transition probability decreasing exponentially with increasing barrier height~\cite{Alkauskas2014}. To make the problem tractable, the multidimensional atomic displacement is projected onto a one-dimensional configuration coordinate $Q$. The total energy is then evaluated along this coordinate using first-principles calculations and fitted within the harmonic approximation, $E = \frac{1}{2}\omega^2 Q^2$, yielding the one-dimensional configuration coordinate (1DCC) diagram shown in Fig. ~\ref{fgr:Figure3}(b)~\cite{stoneham2001theory,Alkauskas2014}, where $\omega$ denotes the effective phonon frequency.\\
We first examine the one-dimensional configuration coordinate (1DCC) diagram for the Sulphur divacancy, shown in Fig. ~\ref{fgr:Figure3}(b). The initial state corresponds to a neutral defect ($V_{S_2}$) with an electron in the conduction band and a hole in the valence band, represented by the minimum of the blue curve at equilibrium coordinate $Q_i$. Upon electron capture, the system transitions to the negatively charged state ($V_{S_2}^{-1}$) with equilibrium configuration $Q_f$, shown by the red curve. This process is accompanied by a substantial lattice distortion of $\Delta Q = 1.68\ \mathrm{amu}^{1/2}\text{\AA}$. The reverse process, corresponding to hole capture and recombination, is represented by the black curve; however, in this work we focus on the electron capture transition from $V_{S_2} + e^- + h$ to $V_{S_2}^{-1} + h$. The capture process is governed by the classical barrier, defined as the energy difference between the minimum of the initial state and the crossing point of the two potential energy surfaces, as indicated in Fig. ~\ref{fgr:Figure3}(b). This barrier depends sensitively on both the lattice distortion and the curvature of the potential energy surfaces. Within the harmonic approximation, the classical capture barrier can be estimated from the crossover point from~\cite{stoneham2001theory,Turiansky2021}
\begin{equation}
E_{b}^{n} = \ \frac{\omega^{2}}{2}\left( \frac{\Delta E}{\omega^{2}\Delta Q} - \frac{\Delta Q}{2} \right)^{2}\
\end{equation}
where, \(\Delta E\) is the position of defect from the conduction band edge i.e. CTL, \(\Delta Q\) is the lattice distortion post charge capture and \(\omega\) is effective frequency of vibration. We note a small barrier \(E_{b}^{n} = 0.23\ eV\) for the case of \(V_{S_{2}}\). For comparison, the prominent C\textsubscript{N} defect in GaN has a barrier \(E_{b}^{n} = 0.73\ eV\) for hole capture~\cite{Alkauskas2014}.\\
Figure ~\ref{fgr:Figure3}(c) shows the configuration coordinate diagram for electron capture at the Sulphur monovacancy. The curvatures of the neutral and charged states are similar, resulting in nearly non-intersecting parabolas. This indicates weak electron–phonon coupling and a tunneling-dominated capture process~\cite{stoneham2001theory}. For the molybdenum vacancy ($V_{\mathrm{Mo}}$), although the lattice distortion ($\Delta Q = 1.04\ \mathrm{amu}^{1/2}\text{\AA}$) is comparable to that of the divacancy, the deeper defect level leads to a large separation between the potential energy surfaces, resulting in a high capture barrier of $E_b^n = 8.40$ eV, as shown in Fig. ~\ref{fgr:Figure3}(d). In contrast, the Sulphur antisite ($S_{\mathrm{Mo}}$), which combines a deep defect level with a relatively small lattice distortion ($\Delta Q = 0.58\ \mathrm{amu}^{1/2}\text{\AA}$), exhibits non-intersecting parabolas (see the SI~\cite{Supplemental}), indicating tunneling dominated carrier capture.\\
Considering full quantum mechanical treatment, we calculate the electron capture coefficient~\cite{Alkauskas2014,Turiansky2021}
\begin{multline}
C = \frac{4\pi^{2}}{h} g V {W_{if}}^{2}
\sum_{m}^{} \Bigg\{
w_{m}
\sum_{n}^{}
\left|
\left\langle
X_{im}\left| \widehat{Q} - Q_{o} \right|X_{fn}
\right\rangle
\right|^{2}
\\
\times
\delta\left(
\Delta E + m\hbar\omega_{i} - n\hbar\omega_{f}
\right)
\Bigg\}
\end{multline}
where $W_{if}$ is the electron–phonon coupling matrix element, $\left\langle X_{im} \left| \hat{Q} - Q_0 \right| X_{fn} \right\rangle$ denotes the phonon overlap matrix element~\cite{Alkauskas2014}, and $\Delta E$ is the charge transition level. 
\begin{figure}[!ht]
  \centering
  \includegraphics[width=1\columnwidth, keepaspectratio]{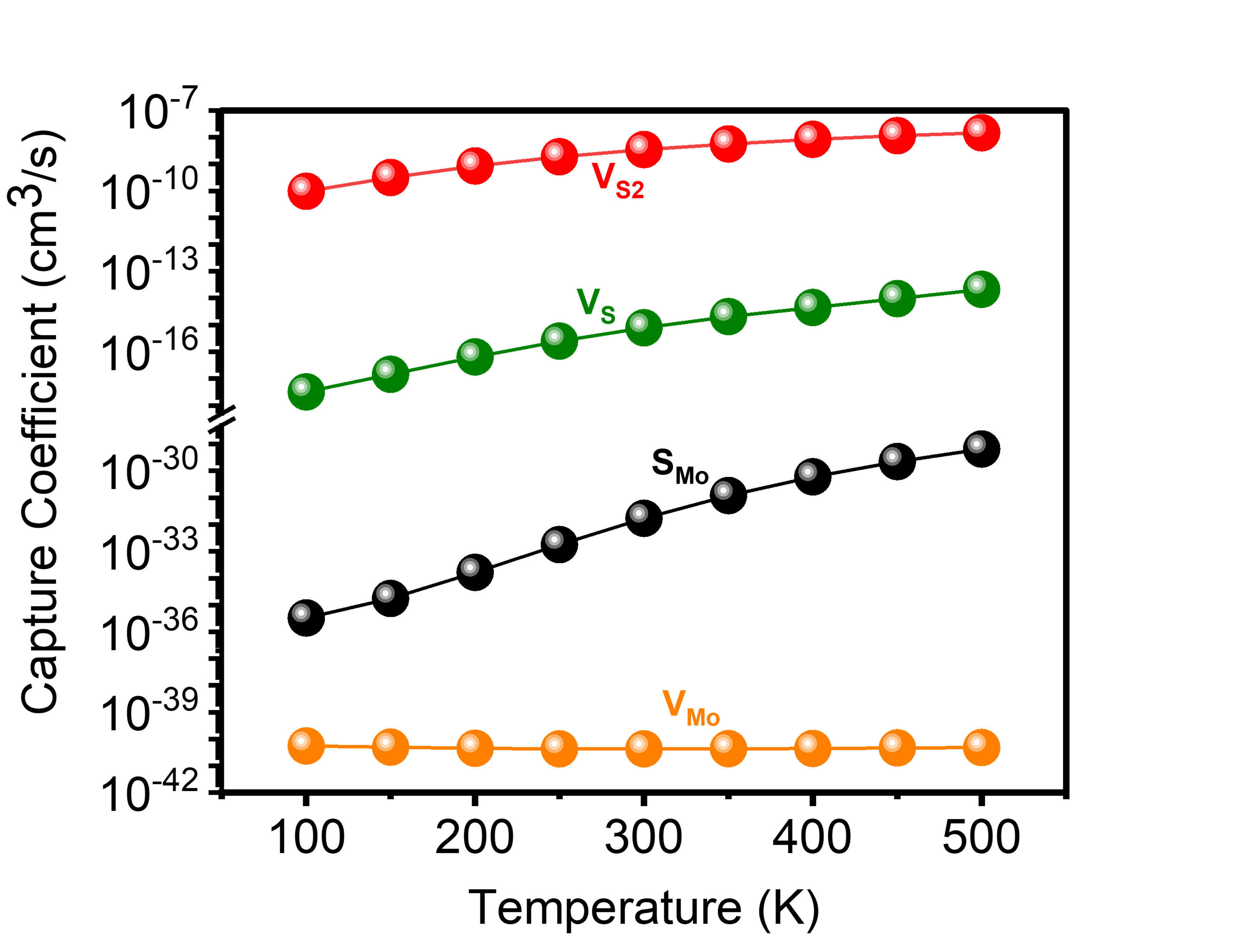}
   \caption{\justifying
Capture coefficient of various intrinsic defects of monolayer MoS$_2$ showing sulphur vacancies as the electron trapper while molybdenum vacancy and sulphur antisite have small capture coefficient. Among all sulphur divacancy has the highest capture coefficient.}
  \label{fgr:Figure4}
  \end{figure}
The quantities $\hbar \omega_i$ and $\hbar \omega_f$ represent the phonon energies of the initial and final states within the harmonic approximation. Here, $g$ is the degeneracy factor, $V$ is the supercell volume, and $w_m$ denotes the thermal occupation of the vibrational state. The coupling matrix element is evaluated within the linear coupling approximation~\cite{Alkauskas2014,Turiansky2021}.
\begin{equation}
W_{if} = \left\langle \left. \ \Psi_{i} \right|\frac{\partial\widehat{h}}{\partial Q}\left| \Psi_{f} \right.\  \right\rangle = {(\varepsilon}_{f} - \varepsilon_{i})\left\langle \left. \ \Psi_{i}(0) \right|\frac{\partial\Psi_{f}(Q)}{\partial Q} \right\rangle\
\end{equation}
\begin{table*}[!t]
\centering
\caption{Defect properties and capture coefficients}
\begin{tabular}{lcccccccc}
\hline\hline
Defects & $\Delta E$ & $\Delta Q$ & $E_b^n$ & $\hbar \omega_f$ & $W_{if}$ & $S$ & $\langle X_{im} | \hat{Q} - Q_0 | X_{fn} \rangle$ & CC    \\
& (eV) & ($\mathrm{amu}^{1/2}\,\mathrm{\AA}$) & (eV) & (meV) & ($\mathrm{eV\,amu^{-1/2}\,\AA^{-1}}$) &  &  & ($\mathrm{cm^3/s}$) \\
\hline\hline
Sulphur monoacancy ($V_S$) 
& 0.26 & 0.48 & 1.68 & 20 & $1.1 \times 10^{-2}$ & 0.5 & $1.1 \times 10^{-14}$ & $7.9 \times 10^{-16}$ \\
Sulphur divacancy ($V_{S_2}$) 
& 0.40 & 1.68 & 0.23 & 19 & $3.3 \times 10^{-2}$ & 7 & $9.9 \times 10^{-6}$ & $3.5 \times 10^{-9}$\\
&  &  &   &  &  &  & & $3.6 \times 10^{-13}$\cite{Ci2020}\\
&  &  &   &  &  &  & & $1.1 \times 10^{-11}$\cite{Ahn2023}\\
Sulphur antisite ($S_{\mathrm{Mo}}$) 
& 1.11 & 0.55 & 5.64 & 35 & $1.2 \times 10^{-3}$ & 1.3 & -- & $1.7 \times 10^{-32}$ \\
Molybdenum vacancy ($V_{\mathrm{Mo}}$) 
& 1.38 & 1.04 & 8.40 & 21 & $7.1 \times 10^{-3}$ & 3 & -- & $4.3 \times 10^{-41}$ \\
\hline
\end{tabular}\label{tbl:tble1}
\end{table*}
The electron–phonon coupling matrix element $W_{if}$ is found to be of the order of $\sim 10^{-2}\ \mathrm{eV},\mathrm{amu}^{-1/2}\text{\AA}^{-1}$ for the Sulphur monovacancy ($V_S$) and divacancy ($V_{S_2}$), whereas for the molybdenum vacancy ($V_{\mathrm{Mo}}$) and Sulphur antisite ($S_{\mathrm{Mo}}$) it is approximately one order of magnitude smaller, $\sim 10^{-3}\ \mathrm{eV},\mathrm{amu}^{-1/2}\text{\AA}^{-1}$. Using the harmonic approximation for phonon overlaps, the resulting electron capture coefficients are shown in Fig. \ref{fgr:Figure4}. \\
The deep defects $V_{\mathrm{Mo}}$ and $S_{\mathrm{Mo}}$ exhibit extremely small capture coefficients of $4.3 \times 10^{-41}\ \mathrm{cm}^3\mathrm{s}^{-1}$ and $1.7 \times 10^{-32}\ \mathrm{cm}^3\mathrm{s}^{-1}$, respectively, at 300 K, consistent with their large capture barriers. In contrast, the Sulphur vacancies show markedly different behavior. The monovacancy ($V_S$) exhibits a capture coefficient of $7.87 \times 10^{-16}\ \mathrm{cm}^3\mathrm{s}^{-1}$ at 300 K. Remarkably, the divacancy ($V_{S_2}$), despite having a slightly deeper defect level, exhibits a capture coefficient of $3.46 \times 10^{-9}\ \mathrm{cm}^3\mathrm{s}^{-1}$, seven orders of magnitude larger. This striking difference demonstrates that carrier capture is not determined solely by the defect level position relative to the band edges, but is strongly governed by the associated lattice relaxation. The calculated capture coefficient for $V_{S_2}$ is in close agreement with experimentally reported values, as summarized in Table~\ref{tbl:tble1}, which lists the key defect parameters and corresponding capture coefficients at 300 K.

\noindent We infer that the molybdenum vacancy ($V_{\mathrm{Mo}}$) and Sulphur antisite ($S_{\mathrm{Mo}}$) exhibit extremely small capture coefficients. This behavior originates from their deep defect levels, which lead to large capture barriers that are inaccessible within the available thermal phonon population. In contrast, both the Sulphur monovacancy ($V_S$) and divacancy ($V_{S_2}$) introduce shallow defect levels. However, their capture coefficients differ dramatically, with values of $7.87 \times 10^{-16}\ \mathrm{cm}^3\mathrm{s}^{-1}$ and $3.46 \times 10^{-9}\ \mathrm{cm}^3\mathrm{s}^{-1}$, respectively, at 300 K. Notably, the divacancy exhibits a capture coefficient seven orders of magnitude larger, despite having a defect level only slightly deeper than that of the monovacancy. To understand this contrast, we note that the capture coefficient within the Born–Oppenheimer framework consists of two independent contributors: (i) the electronic coupling term ($W_{if}$) and (ii) the phonon overlap term $\left\langle X_{im} \left| \hat{Q} - Q_0 \right| X_{fn} \right\rangle$. The electronic coupling $W_{if}$ depends on the overlap between the conduction band and defect wavefunctions. For both $V_S$ and $V_{S_2}$, this quantity is comparable, with values on the order of $\sim 10^{-2}\ \mathrm{eV},\mathrm{amu}^{-1/2}\text{\AA}^{-1}$. The dominant difference arises from the phonon overlap term. While its exact evaluation requires summation over all phonon modes, it can be approximated analytically by a Poisson distribution, characterized by the Huang–Rhys factor ($S$), which represents the strength of electron–phonon coupling, and $\kappa$, the number of phonons involved in the transition~\cite{PhysRev.140.A601,C5CP02093J,Wu2019}.
\begin{equation}
\left\langle X_{im}\left| \widehat{Q} - Q_{o} \right|X_{fn} \right\rangle \propto e^{- S}\frac{S^{\kappa}}{\kappa!}\
\end{equation}
where, \(\kappa = \ \frac{\mathrm{\Delta}E}{h\omega_{f}}\) and \(S\) is given by $S=\frac{1}{2\hslash}(\mathrm{\Delta}Q)^{2}\omega_{f}$. Here, $\Delta Q$ denotes the lattice distortion defined in Eqn. \ref{eqn:eqn1}, and $\omega_f$ is the effective phonon frequency of the defect in the final charge state. Within the Poissonian description of phonon overlap, the transition probability is maximized when the number of phonons involved ($\kappa$) is comparable to the Huang–Rhys factor ($S$), while regimes with $\kappa \gg S$ or $\kappa \ll S$ yield exponentially suppressed overlap. For the Sulphur monovacancy ($V_S$), the Huang–Rhys factor is small ($S < 1$), leading to $\kappa / S \approx 26$ and consequently negligible phonon overlap. In contrast, the Sulphur divacancy ($V_{S_2}$) exhibits $\kappa / S \approx 3$, resulting in significantly enhanced overlap and, therefore, a much larger capture coefficient. As seen from Table~\ref{tbl:tble1}, $V_{S_2}$ exhibits both a larger Huang–Rhys factor and a substantially stronger phonon overlap compared to $V_S$, consistent with its dominant role in carrier capture. For a typical Sulphur monovacancy concentration of $\sim 10^{16}\ \mathrm{cm}^{-3}$, the corresponding trapping rate is on the order of $\sim 1\ \mathrm{s}^{-1}$. This rate is comparable to typical gate-voltage sweep timescales, suggesting that monovacancies can contribute to charge trapping and hysteresis. In contrast, Sulphur divacancies, which are experimentally reported to constitute $\sim$10$\%$ of the total vacancy population~\cite{Zhao2023}, exhibit much faster trapping rates of $\sim 10^{6}\ \mathrm{s}^{-1}$. Such rapid dynamics imply an adiabatic response to gate-voltage sweeps, but can significantly reduce the quantum yield due to efficient nonradiative recombination.
\vspace{-0.4cm}
\section{Conclusion}{\label{sec_4}}
\vspace{-0.45cm}
\noindent In this work, we have quantified electron capture by intrinsic vacancy defects in monolayer MoS$_2$ using first-principles calculations within the nonradiative multiphonon framework. We show that both the Sulphur monovacancy ($V_S$) and divacancy ($V_{S_2}$) introduce shallow defect levels near the conduction band minimum, whereas the molybdenum vacancy ($V_{\mathrm{Mo}}$) and Sulphur antisite ($S_{\mathrm{Mo}}$) form deep states in the band gap. Despite their similar energetic positions, $V_S$ and $V_{S_2}$ exhibit drastically different capture behavior: the divacancy displays a capture coefficient of $\sim 10^{-9}\ \mathrm{cm}^3\mathrm{s}^{-1}$, approximately seven orders of magnitude larger than that of the monovacancy ($\sim 10^{-16}\ \mathrm{cm}^3\mathrm{s}^{-1}$). We demonstrate that this disparity is not controlled by defect level position, but by lattice relaxation, which governs the phonon overlap and hence the multiphonon capture efficiency. From a device perspective, this leads to a clear separation of roles: monovacancies act as slow traps with timescales comparable to gate-voltage sweeps, contributing to hysteresis, whereas divacancies enable fast, efficient nonradiative recombination and suppress quantum yield. In contrast, $V_{\mathrm{Mo}}$ and $S_{\mathrm{Mo}}$, owing to their deep levels and large capture barriers, play a negligible role in electron trapping under typical conditions.\\
More generally, our results establish that carrier capture in two-dimensional semiconductors is governed by the interplay between defect energetics and lattice relaxation. Defects with comparable charge transition levels can exhibit vastly different trapping dynamics when their structural distortions differ. This work provides a quantitative, atomistic framework linking defect structure, electron–phonon coupling, and carrier capture kinetics, and offers clear guidelines for identifying and controlling the defects that limit the performance of two-dimensional electronic and optoelectronic devices.
\vspace{-0.4cm}
\section*{Acknowledgements}
\vspace{-0.4cm}
\noindent K.B acknowledge funding from SERB (grant no: SRG/2021/001783), DRDO
(grant No: JATC-P2QP-14/1335/D(R\&D)/2022, National Quantum Mission
(NQM) under grant no: DST/OTC/NOM/QMD/2024/4 (G)) and Indian Institute
of Technology Delhi Seed grant. S.B acknowledge funding from Anusandhan
National Research Foundation (ANRF), India under Grant No.
CRG/2023/000476. S.S acknowledge funding received from PMRF award by
Government of India. We acknowledge Prof. Saswata Bhattacharya,
Department of Physics, IIT Delhi and Mark Turiansky for useful discussions.
\section*{Competing Interests}
\vspace{-0.4cm}
\noindent The authors declare no financial competing interests.
\vspace{-0.4cm}
\section*{Author Contribution}
\vspace{-0.4cm}
\noindent S.S carried out all the calculations and formal analysis, K.B and S.B supervised the work. All authors have read and approved the final manuscript.

\section*{Data Availability}
\vspace{-0.4 cm}
\noindent The data supporting the findings of this study are available within the paper and its SI~\cite{Supplemental}. The computational input and output files are available from the corresponding author upon reasonable request.
\bibliography{ref_arxiv}

@article{Islam2026,
   author = {Md Mobaidul Islam and Yongin Cho and Anamika Sen and Prashant Bisht and Junoh Shim and Joo-On Oh and Geonyong Park and Antonio Rossi and Hyeongwu Lee and Lin Jiang and Camilla Coletti and Bo-In Park and Heeyeop Chae and SangHoon Shin and Heekyeong Park and Sunkook Kim},
   doi = {10.1038/s41467-026-71986-9},
   issn = {2041-1723},
   issue = {1},
   journal = {Nat. Commun.},
   month = {4},
   pages = {3586},
   title = {Challenges and prospects of 2{D} electronics for future monolithic complementary field-effect transistors},
   volume = {17},
   url = {https://www.nature.com/articles/s41467-026-71986-9},
   year = {2026},
}

@article{das2021transistors,
  title={Transistors based on two-dimensional materials for future integrated circuits},
  author={Das, Saptarshi and Sebastian, Amritanand and Pop, Eric and McClellan, Connor J and Franklin, Aaron D and Grasser, Tibor and Knobloch, Theresia and Illarionov, Yury and Penumatcha, Ashish V and Appenzeller, Joerg and others},
  journal={ Nat Electron.},
  doi ={https://doi.org/10.1038/s41928-021-00670-1},
  volume={4},
  number={11},
  pages={786--799},
  year={2021},
  url = {https://www.nature.com/articles/s41928-021-00670-1},
  publisher={Nature Publishing Group UK London}
}

@article{Ren2025,
   author = {Wencai Ren and Peter Boggild and Joan M Redwing and Konstantin S Novoselov and Luzhao Sun and Yue Qi and Kaicheng Jia and Zhongfan Liu and Oliver Burton and Jack Allen Alexander-Webber and Stephan Hofmann and Yang Cao and Yu Long and Quan-Hong Yang and Dan Li and Soo Ho Choi and Ki Kang Kim and Young Hee Lee and Mian Li and Qing Huang and Yury Gogotsi and Nick Clark and Amy Carl and Roman Gorbachev and Thomas Olsen and Johanna Rosen and Kristian Sommer Thygesen and Prof. Dr. Dmitri K. Efetov and Bjarke S. Jessen and Matthew Yankowitz and Julien Barrier and Roshan Krishna Kumar and Frank Koppens and Hui Deng and Xiaoqin Li and Siyuan Dai and Dmitri Basov and Xinran Wang and Saptarshi Das and Xiangfeng Duan and Zhihao Yu and Markus Borsch and Andrea C Ferrari and Rupert Huber and Mackillo Kira and Fengnian Xia and Xiao Wang and Zhong-Shuai Wu and Xinliang Feng and Patrice Simon and Hui-Ming Cheng and Bilu Liu and Yi Xie and Wanqin Jin and Rahul Raveendran Nair and Yan Xu and Hao-Bin Zhang and Vittorio Pellegrini and Bill Qu and Max Lemme and Ajit Katiyar and Jong-Hyun Ahn and Igor Aharonovich and Mark C Hersam and Stephan Roche and Qilin Hua and Guozhen Shen and Tian-Ling Ren and Chong Min Koo and Nikhil A Koratkar and Robert J Young and Andrew Pollard},
   doi = {10.1088/2053-1583/ae2b82},
   issn = {20531583},
   issue = {2},
   journal = {2D Mater.},
   month = {6},
   pages = {021501},
   publisher = {IOP Publishing},
   title = {The 2{D} Materials Roadmap},
   volume = {13},
   url = {https://iopscience.iop.org/article/10.1088/2053-1583/ae2b82},
   year = {2025},
}

@article{Zhou2013,
   author = {Wu Zhou and Xiaolong Zou and Sina Najmaei and Zheng Liu and Yumeng Shi and Jing Kong and Jun Lou and Pulickel M. Ajayan and Boris I. Yakobson and Juan Carlos Idrobo},
   doi = {10.1021/nl4007479},
   issn = {15306984},
   issue = {6},
   journal = {Nano Lett.},
   month = {6},
   pages = {2615-2622},
   title = {Intrinsic {S}tructural {D}efects in {M}onolayer {M}olybdenum {D}isulfide},
   volume = {13},
   year = {2013},
}

@article{Rasool2015,
   author = {Haider I. Rasool and Colin Ophus and Alex Zettl},
   doi = {10.1002/adma.201500231},
   issn = {15214095},
   issue = {38},
   journal = {Adv. Mater.},
   month = {10},
   pages = {5771-5777},
   publisher = {Wiley-VCH Verlag},
   title = {Atomic Defects in {T}wo {D}imensional {M}aterials},
   volume = {27},
   year = {2015},
}

@article{Karl2025,
   author = {Alexander Karl and Axel Verdianu and Dominic Waldhoer and Theresia Knobloch and Joël Kurzweil and Mina Bahrami and Mohammad Rasool Davoudi and Pedram Khakbaz and Bernhard Stampfer and Seyed Mehdi Sattari-Esfahlan and Yury Illarionov and Aftab Nazir and Changze Liu and Yu Zheng and Lorenzo Pettorosso and Dmitry Polyushkin and Thomas Müller and Saptarshi Das and Xiao Renshaw Wang and Junchuan Tang and Yichi Zhang and Congwei Tan and Ye Li and Hailin Peng and Michael Waltl and Tibor Grasser},
   doi = {10.1038/s41467-025-66210-z},
   issn = {2041-1723},
   issue = {1},
  journal = {Nat. Commun.},
   month = {11},
   pages = {171},
   publisher = {Springer Science and Business Media LLC},
   title = {A standardized approach to characterize hysteresis in 2{D}-materials-based transistors for stability benchmarking and performance projection},
   volume = {17},
   url ={https://www.nature.com/articles/s41467-025-66210-z},
   year = {2025},
}

@article{Kaushik2017,
   author = {Naveen Kaushik and David M.A. Mackenzie and Kartikey Thakar and Natasha Goyal and Bablu Mukherjee and Peter Boggild and Dirch Hjorth Petersen and Saurabh Lodha},
   doi = {10.1038/s41699-017-0038-y},
   issn = {23977132},
   issue = {1},
   journal = {npj 2D Mater Appl .},
   month = {12},
   pages = {34},
   publisher = {Nature Publishing Group},
   title = {Reversible hysteresis inversion in {M}o{S}$_2$ field effect transistors},
   volume = {1},
   year = {2017},
}

@article{PhysRevB.54.11169,
  title = {Efficient iterative schemes for ab initio total-energy calculations using a plane-wave basis set},
  author = {Kresse, G. and Furthm\"uller, J.},
  journal = {Phys. Rev. B},
  volume = {54},
  issue = {16},
  pages = {11169--11186},
  numpages = {0},
  year = {1996},
  month = {Oct},
  publisher = {American Physical Society},
  doi = {10.1103/PhysRevB.54.11169},
  url = {https://link.aps.org/doi/10.1103/PhysRevB.54.11169}
}

@article{Dhosh2025,
   author = {Rittik Ghosh and Alexandros Provias and Alexander Karl and Christoph Wilhelmer and Theresia Knobloch and Mohammad Rasool Davoudi and Seyed Mehdi Sattari-Esfahlan and Dominic Waldhör and Tibor Grasser},
   doi = {10.1016/j.mee.2025.112333},
   issn = {01679317},
   journal = {Microelectronic Engineering},
   month = {9},
   pages ={112333},
   publisher = {Elsevier B.V.},
   title = {Theoretical insights into the impact of border and interface traps on hysteresis in monolayer {M}o{S}$_2$ {F}{E}{T}s},
   volume = {299},
   year = {2025},
}

@article{Park2016,
   author = {Youngseo Park and Hyoung Won Baac and Junseok Heo and Geonwook Yoo},
   doi = {10.1063/1.4942406},
   issn = {00036951},
   issue = {8},
   journal = {Appl. Phys. Lett.},
   month = {2},
   pages ={083102},
   publisher = {American Institute of Physics Inc.},
   title = {Thermally activated trap charges responsible for hysteresis in multilayer {M}o{S}$_2$ field-effect transistors},
   volume = {108},
   year = {2016},
}

@article{Zhao2023,
   author = {Yanfei Zhao and Mukesh Tripathi and Kristiāns Čerņevičs and Ahmet Avsar and Hyun Goo Ji and Juan Francisco Gonzalez Marin and Cheol Yeon Cheon and Zhenyu Wang and Oleg V. Yazyev and Andras Kis},
   doi = {10.1038/s41467-022-35651-1},
   issn = {20411723},
   issue = {1},
   journal = {Nature Communications},
   month = {12},
   pages ={44},
   pmid = {36596799},
   publisher = {Nature Research},
   title = {Electrical spectroscopy of defect states and their hybridization in monolayer {M}o{S}$_2$},
   volume = {14},
   year = {2023},
}

@article{Shu2016,
   author = {Jiapei Shu and Gongtao Wu and Yao Guo and Bo Liu and Xianlong Wei and Qing Chen},
   doi = {10.1039/c5nr07336g},
   issn = {20403372},
   issue = {5},
   journal = {Nanoscale},
   month = {2},
   pages = {3049-3056},
   publisher = {Royal Society of Chemistry},
   title = {The intrinsic origin of hysteresis in {M}o{S}$_2$ field effect transistors},
   volume = {8},
   year = {2016},
}

@article{Ci2020,
   author = {Penghong Ci and Xuezeng Tian and Jun Kang and Anthony Salazar and Kazutaka Eriguchi and Sorren Warkander and Kechao Tang and Jiaman Liu and Yabin Chen and Sefaattin Tongay and Wladek Walukiewicz and Jianwei Miao and Oscar Dubon and Junqiao Wu},
   doi = {10.1038/s41467-020-19247-1},
   issn = {20411723},
   issue = {1},
   journal = {Nat. Commun.},
   month = {12},
   pages = {5373},
   publisher = {Nature Research},
   title = {Chemical trends of deep levels in van der {W}aals semiconductors},
   volume = {11},
   year = {2020},
}

@article{Kim2022,
   author = {Jun Young Kim and Łukasz Gelczuk and Maciej P. Polak and Daria Hlushchenko and Dane Morgan and Robert Kudrawiec and Izabela Szlufarska},
   doi = {10.1038/s41699-022-00350-4},
   issn = {23977132},
   issue = {1},
   journal = {npj 2D Mater Appl .},
   month = {12},
   pages = {75},
   publisher = {Nature Research},
   title = {Experimental and theoretical studies of native deep-level defects in transition metal dichalcogenides},
   volume = {6},
   year = {2022},
}

@article{Alkauskas2014,
   author = {Audrius Alkauskas and Qimin Yan and Chris G. Van De Walle},
   doi = {10.1103/PhysRevB.90.075202},
   issn = {1550235X},
   issue = {7},
   journal = {Phys. Rev. B.},
   month = {8},
   pages = {075202},
   title = {First-principles theory of nonradiative carrier capture via multiphonon emission},
   volume = {90},
   year = {2014},
}

@article{Whalley2021,
   author = {Lucy D. Whalley and Puck Van Gerwen and Jarvist M. Frost and Sunghyun Kim and Samantha N. Hood and Aron Walsh},
   doi = {10.1021/jacs.1c03064},
   issn = {15205126},
   issue = {24},
   journal = {J. Am. Chem. Soc.},
   month = {6},
   pages = {9123-9128},
   pmid = {34102845},
   publisher = {American Chemical Society},
   title = {Giant {H}uang-{R}hys {F}actor for {E}lectron {C}apture by the {I}odine {I}ntersitial in {P}erovskite {S}olar {C}ells},
   volume = {143},
   year = {2021},
}

@article{Turiansky2021,
   author = {Mark E. Turiansky and Audrius Alkauskas and Manuel Engel and Georg Kresse and Darshana Wickramaratne and Jimmy Xuan Shen and Cyrus E. Dreyer and Chris G. Van de Walle},
   doi = {10.1016/j.cpc.2021.108056},
   issn = {00104655},
   journal = {Computer Physics Communications},
   month = {10},
   pages = {108056},
   publisher = {Elsevier B.V.},
   title = {Nonrad: Computing nonradiative capture coefficients from first principles},
   volume = {267},
   year = {2021},
}

@article{Jansen2024,
   author = {Daniel Jansen and Tfyeche Tounsi and Jeison Fischer and Arkady V. Krasheninnikov and Thomas Michely and Hannu Pekka Komsa and Wouter Jolie},
   doi = {10.1103/PhysRevB.109.195430},
   issn = {24699969},
   issue = {19},
   journal = { Phys. Rev. B.},
   month = {5},
   pages = {195430},
   publisher = {American Physical Society},
   title = {Tip-induced creation and {J}ahn-{T}eller distortions of sulfur vacancies in single-layer {M}o{S}$_2$},
   volume = {109},
   year = {2024},
}

@article{Lee2022,
   author = {Yeonghun Lee and Yaoqiao Hu and Xiuyao Lang and Dongwook Kim and Kejun Li and Yuan Ping and Kai Mei C. Fu and Kyeongjae Cho},
   doi = {10.1038/s41467-022-35048-0},
   issn = {20411723},
   issue = {1},
   journal = {Nat. Commun.},
   month = {12},
   pages = {7501},
   pmid = {36473851},
   publisher = {Nature Research},
   title = {Spin-defect qubits in two-dimensional transition metal dichalcogenides operating at telecom wavelengths},
   volume = {13},
   year = {2022},
}

@article{Hong2015,
   author = {Jinhua Hong and Zhixin Hu and Matt Probert and Kun Li and Danhui Lv and Xinan Yang and Lin Gu and Nannan Mao and Qingliang Feng and Liming Xie and Jin Zhang and Dianzhong Wu and Zhiyong Zhang and Chuanhong Jin and Wei Ji and Xixiang Zhang and Jun Yuan and Ze Zhang},
   doi = {10.1038/ncomms7293},
   issn = {20411723},
   journal = {Nature Communications},
   publisher = {Nature Publishing Group},
   title = {Exploring atomic defects in molybdenum disulphide monolayers},
   volume = {6},
   year = {2015},
}

@MISC{Supplemental, 
   author       = "", 
   howpublished = "See Supplemental Information at XXX.", 
   year         = "", 
}

@article{Freysoldt2014,
   author = {Christoph Freysoldt and Blazej Grabowski and Tilmann Hickel and Jörg Neugebauer and Georg Kresse and Anderson Janotti and Chris G. Van De Walle},
   doi = {10.1103/RevModPhys.86.253},
   issn = {15390756},
   issue = {1},
   journal = {Rev. Mod. Phys.},
   month = {3},
   pages = {253-305},
   publisher = {American Physical Society},
   title = {First-principles calculations for point defects in solids},
   volume = {86},
   year = {2014},
}

@article{Ahn2023,
   author = {Byungwook Ahn and Yoonsok Kim and Meeree Kim and Hyang Mi Yu and Jaehun Ahn and Eunji Sim and Hyunjin Ji and Hamza Zad Gul and Keun Soo Kim and Kyuwook Ihm and Hyoyoung Lee and Eun Kyu Kim and Seong Chu Lim},
   doi = {10.1021/acs.nanolett.3c01753},
   issn = {15306992},
   issue = {17},
   journal = {Nano Lett.},
   month = {9},
   pages = {7927-7933},
   pmid = {37647420},
   publisher = {American Chemical Society},
   title = {One-Step {P}assivation of {B}oth {S}ulfur {V}acancies and {S}i{O}$_2$ {I}nterface {T}raps of {M}o{S}$_2$ {D}evice},
   volume = {23},
   year = {2023},
}

@article{PhysRev.140.A601,
  title = {Shapes of {I}mpurity {A}bsorption {B}ands in {S}olids},
  author = {Keil, Thomas H.},
  journal = {Phys. Rev.},
  volume = {140},
  issue = {2A},
  pages = {A601--A617},
  numpages = {0},
  year = {1965},
  month = {Oct},
  publisher = {American Physical Society},
  doi = {10.1103/PhysRev.140.A601},
  url = {https://link.aps.org/doi/10.1103/PhysRev.140.A601}
}

@Article{C5CP02093J,
author ="de Jong, Mathijs and Seijo, Luis and Meijerink, Andries and Rabouw, Freddy T.",
title  ="Resolving the ambiguity in the relation between {S}tokes shift and {H}uang–{R}hys parameter",
journal  ="Phys. Chem. Chem. Phys.",
year  ="2015",
volume  ="17",
issue  ="26",
pages  ="16959-16969",
publisher  ="The Royal Society of Chemistry",
doi  ="10.1039/C5CP02093J",
url  ="http://dx.doi.org/10.1039/C5CP02093J"}

@book{stoneham2001theory,
  title={Theory of Defects in Solids: Electronic Structure of Defects in Insulators and Semiconductors},
  author={Stoneham, A.M.},
  isbn={9780198507802},
  lccn={00126832},
  series={Oxford classic texts in the physical sciences},
  url={https://books.google.co.jp/books?id=jUdrlVC9F0oC},
  year={2001},
  publisher={Clarendon Press}
}

@article{PhysRevMaterials.9.026201,
  title = {Solid-state quantum defects in wide band-gap two-dimensional silica bilayer},
  author = {Lang, Xiuyao and Bergschneider, Matthew and Cho, Kyeongjae},
  journal = {Phys. Rev. Mater.},
  volume = {9},
  issue = {2},
  pages = {026201},
  numpages = {11},
  year = {2025},
  month = {Feb},
  publisher = {American Physical Society},
  doi = {10.1103/PhysRevMaterials.9.026201},
  url = {https://link.aps.org/doi/10.1103/PhysRevMaterials.9.026201}
}

@article{Wu2019,
   author = {Feng Wu and Tyler J. Smart and Junqing Xu and Yuan Ping},
   doi = {10.1103/PhysRevB.100.081407},
   issn = {24699969},
   issue = {8},
   journal = { Phys. Rev. B.},
   month = {8},
   pages = {079901},
   publisher = {American Physical Society},
   title = {Carrier recombination mechanism at defects in wide band gap two-dimensional materials from first principles},
   volume = {100},
   year = {2019},
}

@article{Chatterjee2025,
   author = {Arka Chatterjee and Abhijit Biswas and Addis S Fuhr and Tanguy Terlier and Bobby G Sumpter and Pulickel M Ajayan and Igor Aharonovich and Shengxi Huang},
   doi = {10.1126/sciadv.adv2899},
   issn = {2375},
   issue = {25},
   journal = {Sci.Adv.},
   month = {6},
   pages = {2899},
   title = {Room temperature high purity single photon emission from carbon doped boron nitride thin films},
   volume = {11},
   url ={https://www.science.org/doi/10.1126/sciadv.adv2899},
   year = {2025},
}

@article{Grosso2017,
   author = {Gabriele Grosso and Hyowon Moon and Benjamin Lienhard and Sajid Ali and Dmitri K. Efetov and Marco M. Furchi and Pablo Jarillo-Herrero and Michael J. Ford and Igor Aharonovich and Dirk Englund},
   doi = {10.1038/s41467-017-00810-2},
   issn = {20411723},
   issue = {1},
   journal = {Nat. Commun.},
   month = {12},
   pages = {705},
   publisher = {Nature Publishing Group},
   title = {Tunable and high-purity room temperature single-photon emission from atomic defects in hexagonal boron nitride},
   volume = {8},
   year = {2017},
}

@article{Komsa2015,
   author = {Hannu Pekka Komsa and Arkady V. Krasheninnikov},
   doi = {10.1103/PhysRevB.91.125304},
   issn = {1550235X},
   issue = {12},
   pages ={125304},
   journal = {Phys. Rev. B.},
   month = {3},
   publisher = {American Physical Society},
   title = {Native defects in bulk and monolayer {M}o{S}$_2$ from first principles},
   volume = {91},
   year = {2015},
}

@article{Noh2014,
   author = {Ji Young Noh and Hanchul Kim and Yong Sung Kim},
   doi = {10.1103/PhysRevB.89.205417},
   issn = {1550235X},
   issue = {20},
   journal = {Phys. Rev. B.},
   month = {5},
   pages ={205417},
   publisher = {American Physical Society},
   title = {Stability and electronic structures of native defects in single-layer {M}o{S}$_2$},
   volume = {89},
   year = {2014},
}

@article{Naik2018,
   author = {Mit H. Naik and Manish Jain},
   doi = {10.1103/PhysRevMaterials.2.084002},
   issn = {24759953},
   issue = {8},
   pages = {084002},
   journal = {Phys. Rev. Mater.},
   month = {8},
   publisher = {American Physical Society},
   title = {Substrate screening effects on the quasiparticle band gap and defect charge transition levels in {M}o{S}$_2$},
   volume = {2},
   year = {2018},
}

@article{PhysRevApplied.12.034038,
  title = {Chemical Trend of Transition-Metal Doping in {W}{S}e$_2$},
  author = {Han, Dan and Ming, Wenmei and Xu, Haixuan and Chen, Shiyou and Sun, Deyan and Du, Mao-Hua},
  journal = {Phys. Rev. Appl.},
  volume = {12},
  issue = {3},
  pages = {034038},
  numpages = {23},
  year = {2019},
  month = {Sep},
  publisher = {American Physical Society},
  doi = {10.1103/PhysRevApplied.12.034038},
  url = {https://link.aps.org/doi/10.1103/PhysRevApplied.12.034038}
}

\end{document}